\documentclass{sage}

\usepackage{amsthm,amssymb,natbib,url,amsmath}

\usepackage{epsfig}
\usepackage{graphicx}
\usepackage{amssymb}
\usepackage[noend]{algorithmic}
\usepackage{algorithm}
\usepackage{multirow}
\usepackage{subfigure}
\usepackage{color}

\begin{document}

\title{Performance Analysis and Efficient Execution on Systems with multi-core CPUs, GPUs and MICs}
\author{
George Teodoro$^1$\thanks{Corresponding author;
e-mail: teodoro@unb.br}, Tahsin Kurc$^{2,3}$, Guilherme Andrade$^5$, Jun Kong$^4$, Renato Ferreira$^5$, and Joel Saltz$^{2,3}$\\
{$^1$Department of Computer Science, University of Bras\'ilia, Bras\'ilia, DF, Brazil\\
$^2$Department of Biomedical Informatics, Stony Brook University, Stony Brook, NY, USA \\
$^3$Scientific Data Group, Oak Ridge National Laboratory, Oak Ridge, TN, USA\\
$^4$Department of Biomedical Informatics, Emory University, Atlanta, GA, USA\\
$^5$Department of Computer Science, Federal University of Minas Gerais, Belo Horizonte, MG, Brazil\\
}
}

\maketitle

%\vspace*{-3ex}
\begin{abstract}
We carry out a comparative performance study of multi-core CPUs, GPUs and Intel
Xeon Phi (Many Integrated Core - MIC) with a microscopy image analysis
application. We experimentally evaluate the performance of computing devices on
core operations of the application. We correlate the observed performance with
the characteristics of computing devices and data access patterns, computation
complexities, and parallelization forms of the operations. The results show a
significant variability in the performance of operations with respect to the
device used. The performances of operations with regular data access are
comparable or sometimes better on a MIC than that on a GPU. GPUs are more
efficient than MICs for operations that access data irregularly, because of the
lower bandwidth of the MIC for random data accesses. We propose new
performance-aware scheduling strategies that consider variabilities in
operation speedups. Our scheduling strategies significantly improve application
performance compared to classic strategies in hybrid configurations.

\end{abstract}

\keywords{Hybrid Systems, GPGPU, Intel Xeon Phi, Cooperative Execution, Image Analysis}

\section{Introduction} \label{sec:intro}
Hardware accelerators (co-processors) are becoming increasingly important in 
scientific computing. Through massive parallelism and high-bandwidth memory 
sub-systems, a co-processor can deliver significant performance gains, sometimes at a lower 
energy footprint, in many computationally difficult 
applications~\cite{cudaapps,phi-1,6569806,phi-3,phi-4,DBLP:conf/ipps/SauleC12}. 
As a result, there is a rapid adoption in high-end systems of hybrid configurations equipped 
with multi-core CPUs and one or more co-processors. The 
architectures and execution models of co-processors differ from CPUs and 
among co-processor types, making it challenging for application developers 
to optimize their applications. Evaluating and understanding the performance 
characteristics of co-processors for common computation patterns and operations 
in an application class can help in designing more efficient operations and 
applications, in selecting the most suitable co-processors for an application, and
in developing methods for efficient application execution on hybrid configurations 
with multiple types of co-processors.

Our work targets optimizations for and performance characterizations of  
operations and pipelines that are commonly employed in an important class of 
applications that 
analyze low-dimensional spatial datasets captured by high resolution sensors. 
This class includes applications that analyze and mine large datasets of 
tissue images captured by high resolution digital microscopes; that process data from 
satellites and  ground-based sensors in weather and climate modeling;
that make use of satellite data in large scale biomass monitoring and change analyses;
that analyze seismic surveys in subsurface and reservoir characterization; and
that process wide field survey telescope datasets in
astronomy~\cite{Sharp2006-BMI777,jamia-insilico,508406,Chandola:2011:SGP:2024035.2024041,vatsavai-2011,ParasharMLADW05}.
Input data and data products in these applications are typically
contained in low-dimensional spaces (2D or 3D coordinate system with a temporal
component). Common data processing steps include identification and
segmentation of regions or objects of interest, computation of the texture, 
topology (or structure) and temporal features of segmented objects, classification of 
objects and object groups based on the features, and monitoring and quantifying changes 
over space and time. Table~\ref{tab:op-categories} presents the categories of common
operations in this class of applications and presents examples in microscopy
image analysis, weather prediction, and monitoring and change analysis. These 
operations can be composed into analysis pipelines to address different scientific 
questions.
The data access and processing patterns of the operation categories range 
from local and regular to irregular and global access to data -- see 
Table~\ref{tab:op-data-complexity}~\cite{saltz2013ijhpca}. Local data access patterns 
correspond to accesses to a single
data element or data elements within a small neighborhood in a spatial and
temporal region (e.g., data cleaning and low-level transformations). Regular
access patterns involve sweeps over data elements, while irregular accesses may
involve accesses to data elements in a random manner (e.g., certain types of
object classification algorithms, morphological reconstruction operations in
object segmentation). Some data access patterns may involve generalized
reductions and comparisons (e.g., aggregation) and indexed access (e.g.,
queries for data subsetting and change quantification). 
\begin{table}[t]
\begin{center}
\caption{Operation Categories} 
\begin{footnotesize}
\begin{tabular}{|p{0.17\textwidth}|p{0.24\textwidth}|p{0.24\textwidth}|p{0.24\textwidth}|} 
\hline
Operation Category 			    & Microscopy Image Analysis & Weather Prediction & Monitoring and Change Analysis\\ \hline \hline 
Data Cleaning and Low Level Transformations & 
Color normalization. Thresholding of pixel and regional gray scale values. &
Remove anomalous measurements and convert spectral intensities to values of interest &
Remove unusual readings. Convert signal intensities to color and other values of interest.\\ \hline
Data Subsetting, Filtering, and Subsampling & 
Selection of regions within an image. Thresholding of pixel values.&
Spatial selection/cross match to find portion of a dataset corresponding to a given geographic region. &
Spatial selection/cross match to find portion of a dataset corresponding to a given geographic region. \\ \hline
Object Segmentation & 
Segmentation of nuclei and cells. &
Segmentation of regions with similar land surface temperature. &
Segmentation of buildings, trees, plants, etc.\\ \hline
Feature Computation & 
Compute texture and shape features for each cell. &
Compute texture and shape features for each region. &
Compute texture and shape features for objects.\\ \hline
Spatio-temporal Mapping and Registration & 
Deformable registration of images to anatomical atlas.&
Generation of mosaic of tiles to get complete coverage.&
Registering low and high resolution images corresponding to same regions. \\ \hline 
Classification & 
Clustering of nuclei and/or images into groups.&
Classification of segmented regions.&
Classification of buildings, trees, plants. \\ \hline
Aggregation & 
Aggregation of object features for per image features.&
Time-series calculations on changing land and air conditions. &
Aggregation of labeled buildings, trees, plants into residential, industrial, vegetation areas. \\ \hline 
Change Detection and Comparison & 
Spatial queries to compare segmented nuclei and features within and across images. &
Spatial and temporal queries on classified regions and aggregation to look for changing weather patterns. &
Characterize vegetation changes over time and area.  \\ \hline 
\end{tabular}
\end{footnotesize}
\label{tab:op-categories}
\vspace*{-2ex}
\end{center}
\end{table}

Motivated by the requirements of this class of applications, in this paper we
study the performance impact of GPUs and Intel Xeon Phi (Many Integrated Core -
MIC) co-processors on a set of core operations in the object segmentation and
feature computation categories from Table~\ref{tab:op-categories}. We also
develop and experimentally evaluate several task scheduling strategies for
execution of pipelines of these operations on hybrid systems with multiple
co-processors. We have chosen a microscopy image analysis application, which has been
developed to support large-scale brain tumor
studies~\cite{5518399,jamia-insilico,kong2013machine,kong2013novel,saltz2013large}, as a real application
scenario for performance evaluation. Our choice is driven by the fact that our 
group has extensive experience with this application domain; we have implemented 
methods for object segmentation, feature computation, and classifications and 
analysis pipelines and developed high performance implementations of the methods 
and pipelines.  
In digital pathology high resolution
images obtained from tissue specimens using advanced digital microscopy
instruments are processed to detect and segment nuclei and cells and compute
shape and texture features on the segmented objects. These features are then
mined to investigate the morphology at sub-cellular level of disease and how it
correlates with disease onset and progression and with response to treatment.
In general, execution of a microscopy image analysis application requires a lot
of computing power and memory, because of the computational complexity of
analysis pipelines and the sizes of images -- a three-channel color image
captured by a state-of-the-art scanner can reach 120K$\times$120K pixels in
resolution. In addition, modern scanners are able to collect images quickly and
studies with a few thousand images are becoming common. Adding to the
computational cost is the fact that analysis pipelines may need to be executed
multiple times to tune its parameters and methods for best results or to
determine how sensitive an analysis pipeline is to its input parameters. 
Microscopy image analysis application has many similarities, in terms of data 
access and processing patterns, to other applications 
in our target application class. For example, a change monitoring and analysis 
application will use measurements and images obtained from 
remote sensing instruments (e.g., satellites)~\cite{vatsavai2011,vatsavai2011a,vatsavai2008,upadhyay2008,showalter2001}. 
It will apply similar object detection, feature computation, and classification 
pipelines~\cite{vatsavai2011a,vatsavai2008} to detect and classify different objects and regions on earth 
(e.g., trees, forests, residential areas) -- rather than nuclei and tumor regions 
in a microscopy image analysis application. It may track changes in these areas, like 
a microscopy image analysis application may track changes in distribution of certain 
types of nuclei and tumor regions.   

\begin{table}[t]
\begin{center}
\caption{Data access patterns and computational complexity of common operation categories.} 
\begin{footnotesize}
\begin{tabular}{|p{0.37\textwidth}|p{0.57\textwidth}|}
% {|p{0.17\textwidth}|p{0.24\textwidth}|p{0.24\textwidth}|p{0.24\textwidth}|} 
\hline
Operation Category 			    & Data Access Pattern and Computational Complexity \\ \hline \hline 
Data Cleaning and Low Level Transformations & Mainly local and regular data access patterns. Moderate computational complexity. \\ \hline
Data Subsetting, Filtering, and Subsampling & Local data access patterns as well as indexed access. Low to moderate, mainly data intensive computations. \\ \hline
Object Segmentation & Regular and Irregular, but primarily local, data access patterns. High computational complexity. \\ \hline
Feature Computation &  Regular and Irregular, but primarily local, data access patterns. High computational complexity. \\ \hline
Spatio-temporal Mapping and Registration & Irregular local and global data access patterns. Moderate to high computational complexity. \\ \hline
Classification & Irregular and global data access patterns. High computational complexity.   \\ \hline
Aggregation & Primarily local with a crucial global component for aggregation. Moderate/high computation complexity. \\ \hline
Change Detection and Comparison &  Compute and data-intensive computations.  Mixture of local and global data access patterns as well as indexed access. \\ \hline
\end{tabular}
\end{footnotesize}
\label{tab:op-data-complexity}
\vspace*{-2ex}
\end{center}
\end{table}

Our main contributions can be summarized as follows. {\bf First}, we characterize operations in 
the object segmentation and feature computation categories for the microscopy image analysis application 
according to their data access and computation patterns. {\bf Second}, 
we evaluate the performance of the operations on modern CPUs, GPUs and MIC 
co-processors. We empirically show that the performance gains of the operations vary
significantly according to their data access pattern and parallelization strategies. 
For instance, the performance on a MIC for operations that perform
regular data access is comparable or sometimes better than that on a GPU, but
GPUs are more efficient for algorithms that irregularly access data or rely on
atomic operations. {\bf Third}, we develop new performance-aware scheduling
strategies: PADAS and PAMS, for efficient cooperative execution on hybrid
systems, equipped with CPU and one or multiple accelerators. PAMS attains
better performance than all other policies, including Heterogeneous Earliest
Finish Time (HEFT), for all experiments with multiple accelerators with gains
of at least 1.16$\times$ on top of the competitors. {\bf Finally}, we experimentally 
show that our implementation of the application scales well on distributed memory
systems with different hybrid node configurations: (i)~nodes with CPU and MIC and (ii)~nodes with CPU + MIC + GPU. 

This paper extends our previous work~\cite{Teodoro:2014:CPA:2650283.2650645},
which carried out a performance comparison of image processing operations on
multiple devices, by proposing and evaluating task scheduling strategies for
execution on hybrid configurations with multiple types of co-processors. The
rest of the paper is organized as follows: We describe the motivating
application and the data access and computation characteristics of its core
operations in Section~\ref{sec:app}. Section~\ref{sec:impl} presents the the
programming models and parallelization strategies used to implement the core
operations on GPUs, MICs, and multi-core CPUs.  The deployment of the
application for execution on distributed memory systems and new
performance-aware scheduling policies are presented in
Section~\ref{sec:dist_mic}. We perform an experimental evaluation in
Section~\ref{sec:results}, present related work in Section~\ref{sec:related},
and conclude the paper in Section~\ref{sec:conclusions}.

\section{Example Motivating Application and Core Operations} 
\label{sec:app}
We provide a brief overview of the microscopy image analysis application. Our work is 
presently focused on the development of operations in the object segmentation and feature 
computation categories (or stages), because these are the most expensive stages 
in this application.
We describe the operations and present their data
access and processing patterns. 
\subsection{Microscopy Image Analysis Application}
The microscopy image analysis application is developed primarily to support 
in silico studies of brain cancer~\cite{ieee-insilico,kong2013machine,DBLP:conf/bibm/KongWTCMKPSB13}. These 
studies seek more accurate classifications of tumors by leveraging 
complementary datasets that include high resolution whole tissue slide 
images (WSIs), clinical data and gene expression
data. WSIs are high resolution images with up to 
120K$\times$120K pixels per image. Analysis of a WSI typically
involves: 1)~preprocessing to reduce effects of image acquisition parameters (e.g., 
reduce color intensity variations due to different stainings via
color normalization); 2)~segmentation of microanatomic objects, such as nuclei
and their cytoplasm regions; 3)~feature computation to generate a set of shape and texture 
properties for each object identified
in the segmentation stage; and 4)~clustering of patients (from whom the WSIs have been obtained) 
via machine learning algorithms. The most time consuming stages are 
the segmentation of objects and the computation of features, because the clustering stage 
usually is performed on a significantly reduced dataset, which is realized by 
aggregating and averaging object-level features from each image to compile 
patient-level morphology profiles. 
\begin{figure}[h!]
	\centering
	\includegraphics[width=1\linewidth]{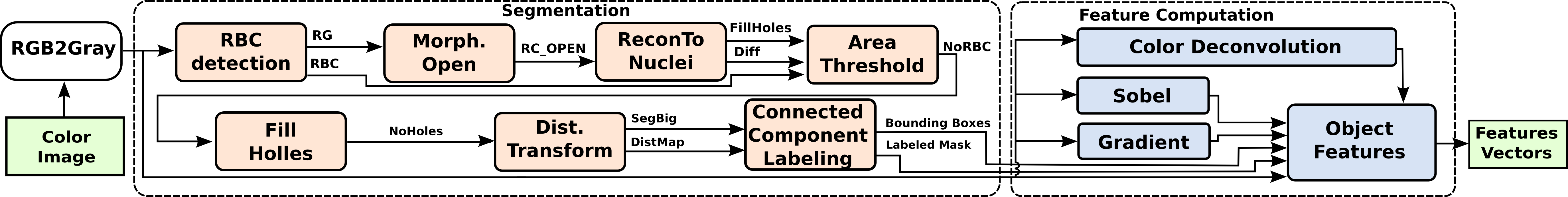}
	\vspace{-5mm}
	\caption{Detailed representation of the Segmentation and Feature
Computation stages of our example microscopy image analysis application. These
stages are composed of several compute intensive finer-grain operations, which
may be naturally represented as a hierarchical computation dataflow.}
	\label{fig:dataflow} 
	\vspace{-2mm} 
\end{figure}

The segmentation and feature computation stages are built by composing several
finer-grain operations as shown in Figure~\ref{fig:dataflow}. The operations in 
\emph{segmentation} include of red blood cells for calculation of potential object 
seeds; morphological open is used to remove noise and fill small holes in seeds; 
morphological reconstruction (ReconToNuclei) identifies candidate objects; area threshold 
filters out objects that are not within a given size range; fill holes fills holes in objects; 
distance transform operation is used to separate overlapping objects; and 
connected component labeling (CCL) creates a labeled mask in which the pixels belonging 
to a segmented object have the same label. The
feature computation stage receives the labeled masks, object bounding boxes,
and the image as input. It executes a set transformations on
the input color image, such as Gradient, Sobel, and Color Deconvolution, which
are then used to derive an array of quantitative attributes for each object 
identified in the segmentation stage. 
\subsection{Data Access and Computation Characteristics of Core Operations} 
\label{sec:core-opts}
The operations in Figure~\ref{fig:dataflow} are categorized in Table~\ref{tab:core-ops} 
according to data access patterns, computation intensity, and the approach for parallel 
execution. This classification is intended to facilitate extrapolation of results from 
our performance evaluations beyond the scope of the example application. We expect that 
operations sharing these characteristics in other applications will have similar 
performance trends. The last column (CPU Source) of the table indicates the source of 
single-core CPU implementations of the respective operations. We have used implementations 
from popular libraries (OpenCV~\cite{opencv_library}) and from other research groups wherever 
possible to ensure that our parallel implementations are compared to efficient baseline 
implementations. 
\begin{table}
{\tiny
\begin{center}
\caption{Segmentation and feature computation operations from the microscopy image analysis application.}
\begin{tabular}{l l l l l l}
\hline
Operations           			& Description				& Data Access Pattern 		& Computation	& Parallelism 	& CPU Source	\\ \hline \hline
\multicolumn{6}{c}{Segmentation Stage} 																\\ \hline
Covert RGB 				& Covert RGB image into			& Regular, multi-channel	& Moderate	& Data		& OpenCV	\\ 
to grayscale				& grayscale intensity image 		& local				& 		&		& 		\\ \hline
Morphological              		& Remove small objects and   		& Regular, neighborhood 	& Low        	& Data		& OpenCV	\\ 
Open 					& fills holes in foreground 		& (13x13 disk)			&               &		&		\\ \hline
Morphological 				& Flood-fill a marker image that 	& Irregular, neighborhood 	& Low        	& IWPP~\cite{Teodoro:2013:Parco}& Vincent~\cite{Vincent93morphologicalgrayscale}\\ 
Reconstruction				& is limited by a mask image.		& (4-/8-connected)		&               &		&		\\ \hline
Area Threshold           		& Remove small or large objects 	& Mixed, neighborhood 		& Low     	& Reduction	& Implemented	\\ \hline
\multirow{2}{*}{FillHolles}             & Fill holes in objects  		& Irregular, neighborhood	& Low		& IWPP~\cite{Teodoro:2013:Parco}		&		\\ 
 					&  using a flood-fill			& (4-/8-connected)		&       	&		& Vincent~\cite{Vincent93morphologicalgrayscale}	\\ \hline
Distance  				& Compute distance to 			& Irregular, neighborhood	& Moderate 	& IWPP~\cite{Teodoro:2013:Parco}	        & Implemented	\\ 
Transform				& closest background pixel 		& (8-connected)			&               &		&		\\ 
					& for each foreground pixel		& (8-connected)			&               &		&		\\ \hline
Component           			& Label with the same value  		& Irregular, global		& Low 		& Union-find	& Oliveira~\cite{Oliveira10}	\\ 
Labeling				& each component of the mask		&				&  		&		&		\\ \hline \hline
\multicolumn{6}{c}{Feature Computation Stage} 															\\ \hline
Color Decon-				& Separate multi-stained 		& Regular, multi-channel 	& Moderate	& Data		& Implemented	\\ 
volution~\cite{Ruifrok_2001}		& image into different channels		& local				&		&		&		\\ \hline
\multirow{2}{*}{Pixel Statistics} 	& Compute intensity stats (mean,  	& Regular, access a set		& High		& Object 	& Implemented	\\ 
					& median, max, etc) per object 		& of bounding-boxed areas	&   		&		&		\\ \hline
Gradient 				& Calculate x,y gradient		& Regular, neighborhood 	& High		& Object	& Implemented	\\ 
Statistics				& and derive per object stats		& and bounding-boxed areas	&		&		& OpenCV(Grad)	\\ \hline	
\multirow{2}{*}{Sobel Edge} 		& Compute Sobel and derive 		& Regular, access a set		& High		& Object	& Implemented	\\ 
					& per object stags 			& of bounding-boxed areas	&   		&		& OpenCV(Sobel)	\\
\hline
\end{tabular}
%\vspace*{-2ex}
\label{tab:core-ops}
\vspace*{-6ex}
\end{center}
}
\end{table}

The operations are first classified according to their data access patterns: 1)~regular 
operations access and process contiguous elements or chunks of data in the domain space, 
such as scanning pixels in an input image; 2)~irregular operations access data 
elements irregularly distributed in the data domain, and data elements to be 
accessed are determined at runtime as the computation progresses and dependencies 
are resolved. Within each of these classes, the operations can further be categorized 
based on how the value of a data element is computed or updated: 1)~{\em local}
when the computation involves only the value of the elements being processed;
2)~{\em multi-channel local} if multiple data elements with the same spatial
location across multiple layers of the domain are accessed (e.g, the same index in
multiple image channels); 3)~{\em neighborhood} if computation on a data element involves 
values from data elements in a spatial neighborhood of the data element. The neighborhood could be
defined using a structure centered at the element being operated on. Examples of
these structures are 4-/8-connected components or discs (defined with a
radius from the current element); and 4)~{\em areas within bounding-boxes} if computations 
on data elements involve data elements within a region defined by a bounding-box. This type 
of computation is primarily a characteristic of the operations in the feature computation 
stage that process data in regions defined by the minimum bounding boxes of segmented 
objects. 

Implementations of the operations for parallel execution can employ a variety of 
parallelism strategies: 1)~Data parallelism;
2)~Object parallelism; 3)~MapReduce~\cite{dean04mapreduce} or generalized
reduction; 4)~Irregular wavefront propagation pattern
(IWPP)~\cite{Teodoro:2013:Parco}; and 
5)~Union-find~\cite{Tarjan:1975:EGB:321879.321884}. Data parallelism is suitable 
for operations that carry out computation of elements in the data domain
independently. Object parallelism corresponds to parallel and independent
computation of data objects. The Generalized Reduction or MapReduce pattern is 
used in the "Area Threshold" operation. The first step of Area Threshold 
computes a mapping of foreground pixels based on their labels -- the label of a 
pixel indicates which candidate object it belongs to; pixels that are not part of 
a candidate object are labeled with 0 (zero). The second step executes a reduction 
to count the number of pixels with the same label value. The count indicates the 
area of the corresponding candidate object. In the last step, objects whose areas 
are not within a given range are removed from further consideration. 

The IWPP pattern executes independent wavefront propagations that may start
from multiple source elements in the domain. Elements located in the front of
these waves, called active elements, serve as sources of propagation to their
neighbor elements. The value of an element may be modified by a propagation
from neighbor active elements in a wavefront. The IWPP pattern is presented in
Algorithm~\ref{alg:genericRecon}. A set of (active) elements from a
multi-dimensional grid space ($D$) is selected to initialize the wavefront
($S$). The algorithm then enters the wavefront propagation phase in which an
element ($e_i$) from S is removed for computation. That element will attempt to
propagate its value to neighboring elements ($N_G$) in the $G$ structure.
Whenever the propagation condition between $e_i$ and each neighbor ($e_j \in
Q$) is satisfied, $e_j$ receives the propagation and is added to the set of
active elements in the wavefront. Since $e_j$ has its value modified, it may
now be able to change the value of one of its neighbors. This process is
similar to the process that takes place in flood-fill algorithms. The wavefront
propagations are repeated until the set $S$ is empty.

Because the management of the wavefront is a core part of the IWPP algorithm, the 
algorithm's performance (in both sequential execution and parallel execution) depends 
on using an efficient means (e.g., using a queue or a set) to track active elements 
and, as a consequence, to avoid computations by data elements that do not contribute 
to the output.  
\begin{algorithm}
\begin{center}
\begin{small}
\caption{Irregular Wavefront Propagation Pattern (IWPP).}
\label{alg:genericRecon}
\begin{algorithmic}[1]
\STATE $D \leftarrow$ data elements in a multi-dimensional space
\STATE \{{\bf Initialization Phase}\}
\STATE $S \leftarrow$ select active elements from $D$
\STATE \{{\bf Wavefront Propagation Phase}\}
\WHILE{$S \neq \emptyset$}
	\STATE Extract $e_i$ from $S$
	\STATE $Q \leftarrow$ $N_G(e_i)$
	\WHILE{$Q \neq \emptyset$}
		\STATE Extract $e_j$ from $Q$
		\IF{$PropagationCondition$($D(e_i)$,$D(e_j)$) $=$ true}
			\STATE $D(e_j) \leftarrow$ $Update$($D(e_i)$)
			\STATE Insert $e_j$ into $S$
		\ENDIF
	\ENDWHILE
\ENDWHILE 
\end{algorithmic}
\end{small}
\end{center}
\end{algorithm}

The \emph{union-find pattern}~\cite{Tarjan:1975:EGB:321879.321884} is used for
manipulating disjoint-set data structures that may be merged efficiently.
Rooted trees are used to represent sets; an arbitrary member of a set is
used as its representative or root of the tree. The non-root element point to its
parent and root points to itself. The union-find is implemented using the following
four operations: (1)
{\tt MakeSet(x)}: creates an elementary set containing only x, which is not a member 
of any other set;
(2) {\tt FindRoot(x)}: determines the root of the set in which x is store;
(3) {\tt Union(x,y)}: merges the sets containing elements x and y;
(4) {\tt Criterion(x,y)}: determines if x and y belong to the same set.
The connected components labeling (CCL) operation implements the union-find
pattern.  The CCL first creates a forest in which each element (pixel) from the
input image is an independent tree. It iteratively merges trees from adjacent
elements in the data domain such that one tree becomes a branch in another
tree. The criterion for merging trees is that the neighbor elements must be
foreground pixels in the original image. When merging two trees (Union), the
label values of the roots of the two trees are compared, and the root with the
smaller value is selected as the root of the merged tree.  After this process
has been done for all the pixels in the image, each connected component is
assigned to a single tree.

\section{Implementations of Operations on multi-core CPU, MIC and GPU} \label{sec:impl}
%
% \subsection{Architectures and Programming Models} \label{sec:arch}

Programming tools and languages for code development for a MIC are
similar to those used for CPUs. This is a significant advantage as compared to
GPUs. It alleviates code migration overheads. The MIC
supports several parallel programming languages and models, e.g., OpenMP, Intel
Threading Building Blocks (TBB), and Intel Cilk Plus.  In this work, we have
implemented the parallel versions of the operations on the MIC and CPU using
OpenMP, because it is a well known parallelization framework that only requires 
annotation of the code. The implementations for the GPU are done with  
CUDA\footnote{http://nvidia.com/cuda/.}. The MIC supports two execution modes:
native and offload. In the native mode the application runs entirely within the
co-processor, because MICs run a specialized Linux kernel that
provides the necessary services and interfaces to applications. The offload
mode allows for the CPU to execute regions of the application code with a MIC.
These regions are defined using pragma tags and include
directives for transferring data. The offload mode also supports conditional
offload directives, which can be used to decide at runtime whether a region
should be offloaded to the co-processor or should be executed on the CPU.  This
feature is used in our dynamic task assignment strategy to use the CPU and MIC 
cooperatively for application execution.   
We present the details of the MIC and GPU implementations only, because the CPU
runs the same code as the MIC.

%\subsection{Parallel Implementation of Operations} \label{sec:opt-impl}

The \emph{data parallel} operations are
trivial to implement, since computing threads may be assigned for independent
computation of elements from the input data domain. However, we had to analyze
the results of the auto vectorization performed by the compiler for the MIC,
since it could not vectorize some of the loops with complex pointer
manipulations. We annotated those codes with (\#pragma simd) directives to guide 
the vectorization where appropriate.

The implementation of an operation with the \emph{IWPP pattern} makes use of efficient 
parallel containers to store the wavefront elements. The
parallel computation of elements in the wavefront requires those elements be
atomically updated, as multiple elements may concurrently update an 
element $e_j$. We have developed a complex hierarchical parallel queue to store wavefront
elements for the GPU implementation~\cite{Teodoro:2013:Parco}. 
The parallel queue exploits the multiple
GPU memory levels and is implemented in a thread block basis, such that each
block of threads has an independent instance of the queue to avoid
synchronization among blocks. The implementation of the IWPP for the MIC utilizes
the standard C++ queue data structure as the container. The IWPP code 
instantiates one copy of this container per computing thread. Each computing 
thread independently
carries out propagations of a subset of wavefront elements. Atomic
operations are used to update memory during a propagation to avoid race
conditions. As a result, the MIC vectorization is not possible since
vector atomic instructions are not supported.

The \emph{MapReduce-style pattern} is employed in object area calculations. The
MIC and GPU implementations use a vector with an entry per object to accumulate
its area. Threads concurrently scan pixels in the input data domain to
atomically increment the corresponding entry in the reduction vector. Because
the number of objects may be very high, it is not feasible to create a copy of
this vector per thread and eliminate the need for atomic instructions.

In the \emph{Union-find pattern} a forest is created in the input image, such
that each pixel stores its neighbor parent pixel or itself when it is a tree
root. For the parallelization of this pattern, the input image data is divided
into tiles that may be independently processed in parallel. A second phase is
executed to merge trees that cross tile boundaries. The MIC implementation
assigns a single tile per thread, which eliminates the need for atomic
instructions in the first phase. The GPU implementation, on the other hand,
computes each tile using a thread block. Since the threads computing a tile are
in the same block, they can take advantage of the fast shared-memory atomic
instructions. The implementation of the second phase of Union-find is similar
for both the MIC and the GPU. It uses atomic updates to guarantee consistency
during tree merges across tile boundaries.
The \emph{object parallel} operations are concurrently executed with each
instance processing a segmented object. A single thread in the MIC or a block
of threads in the GPU is assigned for the computation of features related to
each object. All the operations with this type of parallelism are fully
vectorized.

\section{Execution on Hybrid Cluster Systems} \label{sec:dist_mic}
Execution of the microscopy image analysis application on a hybrid cluster system 
is supported by a runtime system designed for dataflow applications and developed 
in our previous work~\cite{Teodoro-IPDPS2013,Teodoro-IPDPS2012}. This runtime system 
expresses an application as a hierarchical dataflow graph with multiple levels, which
allows for a computation task in one level to be described as another dataflow
of finer-grained operations. The hierarchical representation allows for more flexibility 
in scheduling of application stages and operations. For example, different scheduling 
strategies can be used in each level.   
The microscopy image analysis application is represented 
in this runtime system as a two-level computation graph. The first
level consists of coarse-grain computing stages: segmentation and feature
computation. Each of these stages is represented as a graph of finer-grain
operations which forms the second level. This is a natural representation of the 
application, which, as presented in Figure~\ref{fig:dataflow}, has a hierarchical 
organization of computation.
\subsection{Execution on a Distributed-memory Computation Cluster}
\begin{figure}[h!]
\centering
    \includegraphics[width=0.96\linewidth]{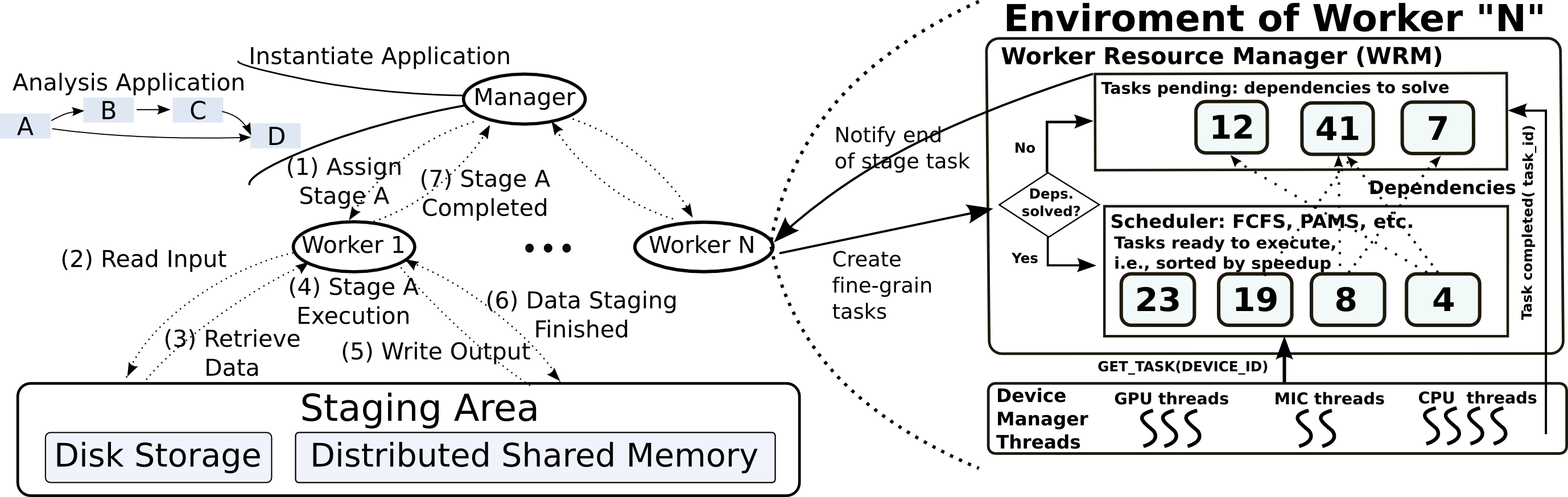}
\caption{Runtime system architecture}
\label{fig:runtime}
\end{figure}

Our parallel execution strategy is based on a Manager-Worker model that
combines a bag-of-tasks style execution with coarse-grain dataflow and makes
use of function variants (see Figure~\ref{fig:runtime}) -- a function variant
represents multiple implementations of a function with the same signature; in
our case, the function variants of an operation are the CPU, GPU, and MIC
implementations. Each image in a dataset is partitioned and processed in tiles.
Image tiles can be processed concurrently. The parallel execution strategy
assigns the first level (coarse-grain) stage tasks and image tiles to the nodes
of the cluster in a demand-driven fashion. Within each node, it employs another
task scheduling method (see Section~\ref{sec:sched}) to map the second level
(finer-grain) tasks to computing devices. 

One Manager process is instantiated on the head node and each computation node
is designated as a Worker. The Manager creates tasks of the form (input image
tile, processing stage), where {\em processing stage} is either the
segmentation stage or the feature computation.  Each of these tasks is referred
to as a stage task. The Manager also builds the dependencies between stage task
instances to enforce correct execution. The stage tasks are scheduled to the
Workers using a demand-driven approach, and a Worker may request and compute
multiple stage tasks in parallel to use all the computing devices on a node. 
A local Worker Resource Manager (WRM) is created on each Worker to 
manage CPU cores and co-processors (GPUs or MICs) available. The WRM
creates a device manager thread to control each accelerator or CPU core.  When
the Worker receives a stage task, it creates fine-grain tasks that implement
the stage task and dispatch them for execution with the WRM.  The assignment of
fine-grain tasks to computing devices in a node is carried out in a
demand-driven basis. Whenever a device manager thread becomes idle, it
requests a task for execution with the WRM. The WRM then selects a task among
those with dependencies resolved using either a First Come, First Served (FCFS)
strategy or one of the performance-aware scheduling algorithms proposed in
Section~\ref{sec:sched}.  After all fine-grain tasks of a given
coarse-grain stage have been executed, the WRM notifies the Manager process and
requests more coarse-grain tasks. This process continues until
all coarse-grain or stage tasks have been processed.

\subsection{Performance-Aware Task Scheduling} \label{sec:sched}
We propose new scheduling strategies for execution of tasks on a node with
multi-core CPUs and multiple co-processor types or models. In our earlier
work~\cite{Teodoro:2014:CPA:2650283.2650645}, we used a Performance-Aware Task
Scheduling (PATS) strategy to efficiently assign computation for cooperative
execution on a node with CPUs and one type or model of co-processor. PATS
assigns tasks to CPU cores or co-processors based on the load of the computing
devices and an estimate of each task's speedup on the co-processor. When a
co-processor requests a task, PATS assigns the task with estimated higher
speedup to the co-processor. If the device available for computation is a CPU,
the task with the estimated lower speedup is chosen. PATS was built targeting
systems with CPUs and a single type of co-processor. It is not able to handle
scheduling in environments with CPUs, GPUs, and MICs. This limitation of PATS
motivated the development of new performance-aware scheduling policies
presented in the remaining of this section. In these policies, assignment of
tasks to devices is performed on a demand-driven basis as computing devices
become available. 

\begin{figure}[]
\centering
\hspace{0.35cm}
\mbox{
    \subfigure[PADAS]{
    \includegraphics[width=0.46\linewidth]{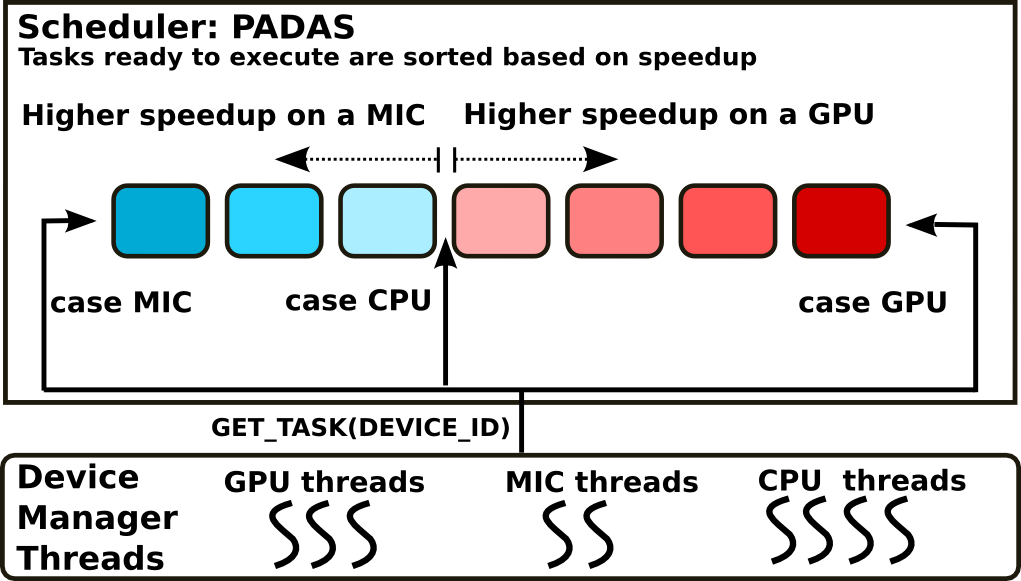}}
}
%\hspace{0.25cm}
\mbox{
    \subfigure[PAMS]{
    \includegraphics[width=0.46\linewidth]{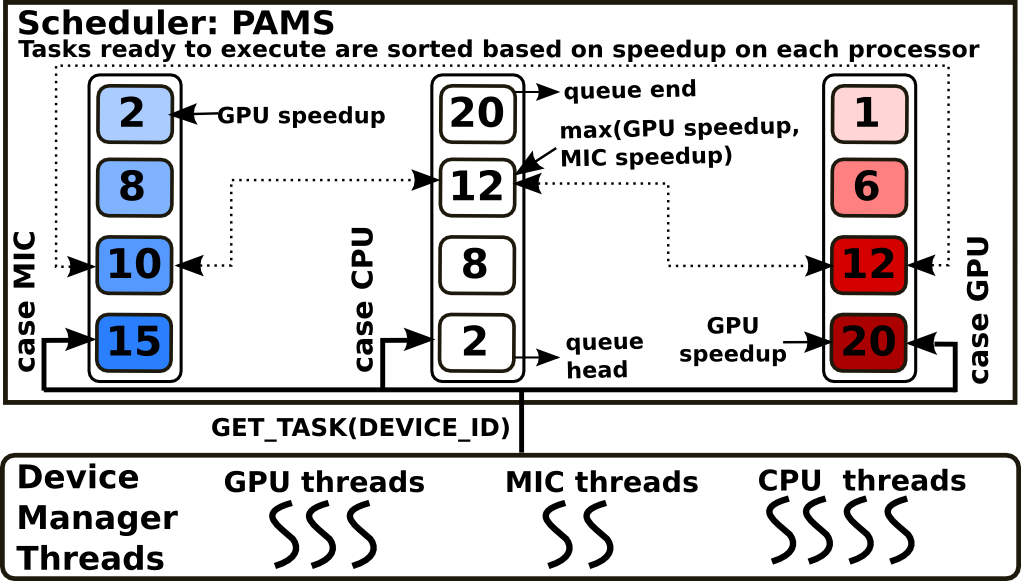}}
}
\vspace{-1.2mm}
\caption{Proposed Performance-Aware Scheduling Strategies}
\vspace{-0.6cm}
\label{graf:scheds}
\end{figure}

\paragraph{{\bf Performance-Aware Dequeue Adapted Scheduler (PADAS):}} is
implemented using a single queue to hold tasks ready for execution. This queue
maintains tasks in sorted order based on the expected speedup of each task on
the co-processors. Insertion of tasks in this queue may start from the head or
tail of the queue. If the expected acceleration of a task on a MIC is higher
than that on a GPU, the task is inserted into the queue from the head. The
task is placed in the queue before the first task with a smaller speedup on the MIC.
If the expected speedup on a GPU of the task is higher than that on the MIC,
the tasks is inserted from the end of the queue. The task is placed in the
queue before the first task with a smaller expected speedup on the GPU. The reasoning
with this strategy is to keep tasks that benefit the most from the MIC in one
end of the queue and tasks more suitable for the GPU in the other. The use of a
single queue leads to reduced overheads and allows for quick retrieval of the
most appropriate task to the type of processor available.  When a MIC requests
a task, PADAS will retrieve the first task from the head of the queue.  When a
GPU is available, the task in the tail is chosen. The CPU cores consume tasks
from the middle point between tasks inserted from the head (MIC) and end (GPU)
of the queue (see Figure~\ref{graf:scheds}).  

\paragraph{{\bf Performance-Aware Multiqueue Scheduler (PAMS):}} uses multiple
queues; one queue for each type of computing device. The task queues are
maintained sorted according to the expected speedup of tasks on the target
device. The GPU and MIC queues are sorted in descending order, whereas the CPU
queue uses the $max(GPU\ Speedup, MIC\ Speedup)$ as a metric to order its tasks
in ascending order. This strategy aims to use CPU cores for tasks that are not
efficiently executed by co-processors.  Task insertions are performed in all
queues. A pointer to an inserted task exists in each queue. The pointer is used
to efficiently delete a task from the other queues once it is selected for
execution from a queue.  The process of scheduling a task for computation is
simple. When a device becomes idle, the task on the head of its queue is
selected for execution, and the pointers built during the insertion are used to
quickly delete that task from the other queues.  Figure~\ref{graf:scheds}
presents a scheme of this scheduling strategy.

The performance-aware scheduling strategies we proposed only rely speedups to
maintain order of tasks in queues. Therefore, if errors in estimate of speedups
are not sufficient to change order of tasks, the performance of our policies
will not be affected. In fact, our performance evaluation shows that scheduling
strategies based on execution times, such as Heterogeneous Earliest Finish Time
(HEFT), are more sensitive to errors in estimates. As a consequence, they
require more accurate input to compute an efficient mapping of tasks to
processors.

Task scheduling in hybrid machines has been investigated in a number of
projects~\cite{mars,merge,qilin09luk,1616864,Diamos:2008:HEM:1383422.1383447.Harmony,6061070,ravi2010compiler,6152715}.
Scheduling strategies developed in prior work partition tasks among computing devices (i.e., CPUs and
GPUs) for applications in which application tasks/operations attain the same level of speedup
on an accelerator. However, applications that apply multiple, different
transformations to input data are composed of operations which accelerate 
at different speeds on an accelerator. The scheduling strategies we
proposed take these speedup variations into account in order to better use hybrid
system, whereas the other approaches will commonly only try to minimize load
imbalance among computing devices.

\section{Experimental Performance Evaluation} \label{sec:results}
We conducted a series of experiments to compare the performance impact of 
co-processors on image analysis operations and to evaluate the 
scheduling algorithms for efficient execution on a computational cluster with 
co-processors. We used a distributed-memory Linux cluster, called Stampede, for this 
purpose. Stampede is one of the high performance computing resources provided by 
XSEDE\footnote{https://www.xsede.org/home}; the machine is hosted at Texas Advanced 
Computing Center\footnote{https://portal.xsede.org/tacc-stampede}. 
Each compute node of Stampede has dual socket Intel Xeon E5-2680 processors, an Intel 
Xeon Phi SE10P co-processor and 32GB main memory. The cluster also has 
128 nodes each with one SE10P and one NVIDIA K20 GPU. 
The hardware details of the Intel Phi and GPU used in our evaluation
are presented in Table~\ref{tab:devices}. All the nodes in the cluster 
are connected via Mellanox FDR InfiniBand switches. 

In all of the experiments the MIC based codes were executed using the offload mode. One MIC computing 
core was reserved for running the offload daemon. The MIC and multi-core CPU codes for the 
image analysis operations were implemented using C++ and OpenMP, while the GPU codes were 
developed using CUDA. We measured processing performance using a 
set of images which had been collected from The Cancer Genome Atlas repository for brain 
tumor studies~\cite{kong2013machine}. Each image is partitioned into 4K$\times$4K tiles, which 
can be processed concurrently for image analysis.

While we used a microscopy image analysis application for performance evaluation, we believe our 
findings will be applicable in other applications that have data access and processing 
characteristics similar to those of the microscopy image analysis application. These characteristics 
can be summarized as follows: (1) spatial data domain is partitioned into chunks (or tiles), and 
tiles are processed concurrently; (2) processing of each tile is expressed as a 
pipeline of operations; (3) the pipeline operations have a variety 
of data access and processing patterns (see segmentation and feature computation categories in 
Tables~\ref{tab:op-categories}~and~\ref{tab:op-data-complexity}), which lead to variable speedups on 
co-processors; (4) the variability in speedups also arise from the fact that some pipeline operations 
are data intensive while others are computationally more expensive. We observed similar 
processing characteristics in applications that analyze satellite imagery and that process telescope 
datasets. 

\begin{table}
\begin{center}
\caption{Hardware specifications of the MIC and GPU used in the performance experiments.} 
\begin{footnotesize}
\begin{tabular}{p{0.14\textwidth}p{0.2\textwidth}p{0.2\textwidth}} 
\hline
		& Intel Xeon Phi SE10P 	& NVIDIA K20 GPU  	\\ \hline \hline 
\# of cores	& 61 	 		& 2496		 	\\ \hline
Clock		& 1.09~GHz		& 0.706~GHz	\\ \hline
L1 Cache	& 32~KB			& 64~KB \\ \hline
L2 Cache	& 512~KB		& 768~KB\\ \hline
Memory		& 8~GB			& 5~GB\\ \hline
Bandwidth	& 352~GB/s		& 205~GB/s \\ \hline
\end{tabular}
\end{footnotesize}
\label{tab:devices}
\vspace*{-2ex}
\end{center}
\end{table}

\subsection{Performance and Scalability of Operations on MIC} \label{sec:opts-scalability}
\begin{figure}[htb!]
\begin{center}
        \includegraphics[width=0.96\textwidth]{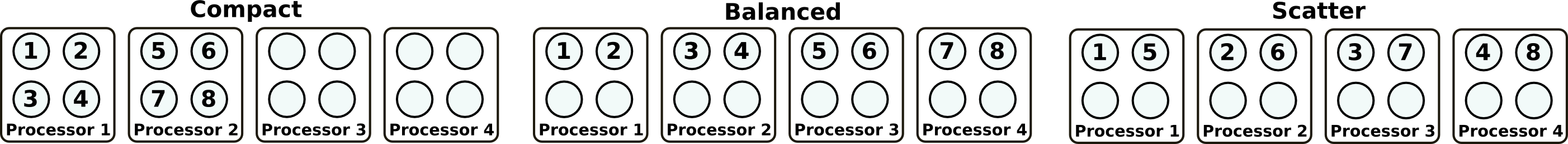}
\vspace*{-1ex}
\caption{Thread affinity strategies. Example mappings of 8 threads to physical 
cores on a 4-way hyper-threaded device.}
\vspace*{-3ex}
\label{fig:affinity}
\end{center}
\end{figure}

Thread affinity (i.e., how process threads are mapped to the physical cores of a MIC) 
can affect the performance of operations running on a MIC. In this set of experiments 
we evaluate the performance impact of three thread affinity strategies 
(Figure~\ref{fig:affinity}) on the core operations described in Section~\ref{sec:app}. 
The first strategy, called 
{\em Compact}, binds threads to the next free thread context. That is, 
all four contexts in a physical MIC core are used before threads are placed
in the contexts of another core. The second strategy, called {\em Balanced}, 
allocates threads to new cores before all the contexts in the same core are 
used.  Threads are balanced among cores and subsequent thread IDs are assigned to
neighbor contexts. The last strategy, {\em Scatter}, also balances the computation, but
threads with subsequent IDs are assigned to different physical cores.

The experimental evaluation was performed using two operations with different 
data access and computation patterns: (1)~Morphological Open has a regular data
access pattern and low computation intensity; (2)~Distance Transform performs
irregular data accesses and has moderate computation intensity. The codes 
employed OpenMP static loop scheduling, because it led to better performance. 

\begin{figure}[htb!]
\begin{center}
	\subfigure[Morphological Open]{\label{fig:speedup-open}
        \includegraphics[width=0.46\textwidth]{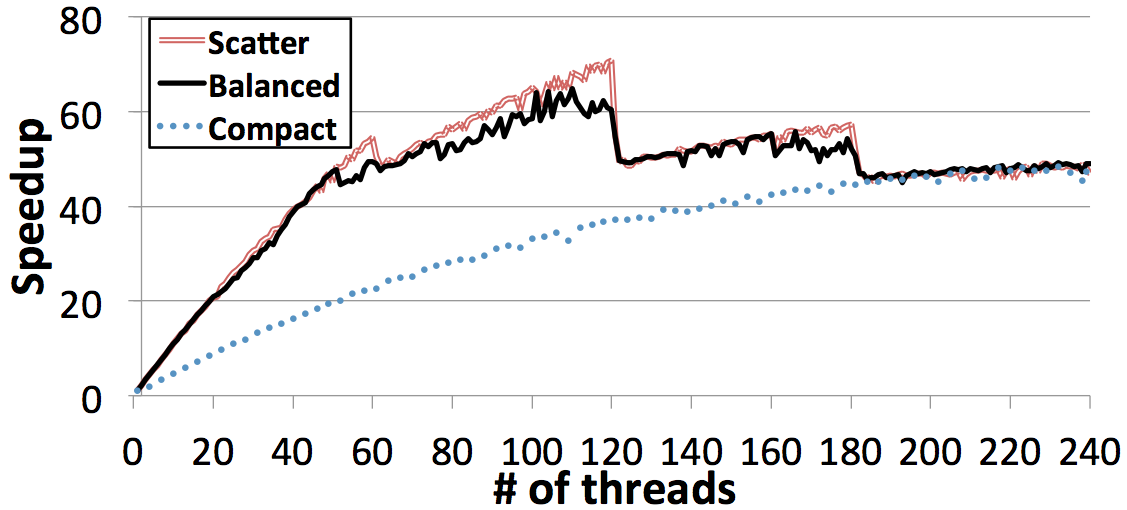}}
%	\subfigure[]{\label{fig:speedup-grad}
 %       \includegraphics[width=0.32\textwidth]{results/1-node/speedup/grad}}
	\subfigure[Distance Transform]{\label{fig:speedup-dist}
        \includegraphics[width=0.44\textwidth]{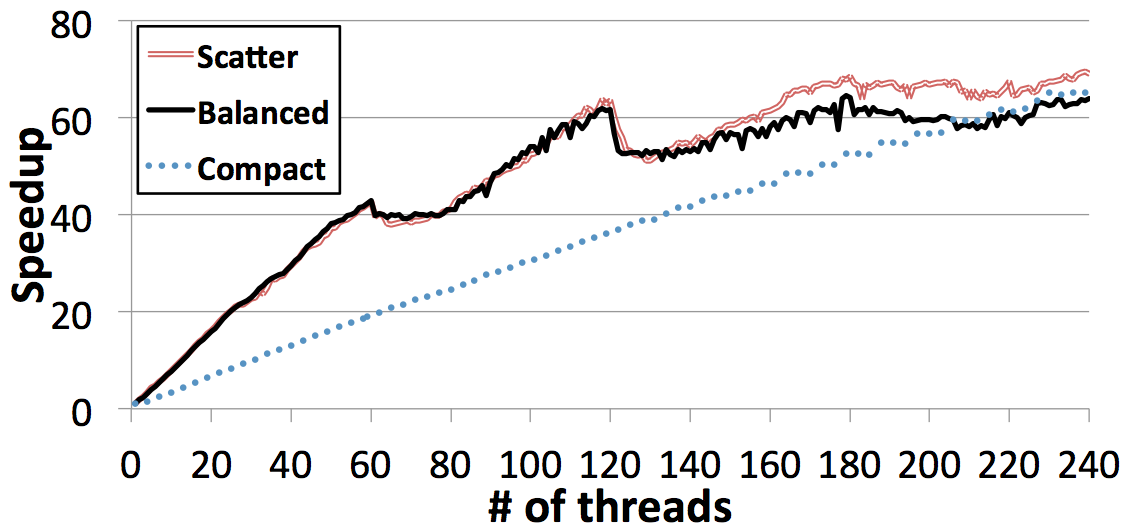}}
\vspace*{-1ex}
\caption{Scalability of Morphological Open and Distance Transform under three 
thread affinity strategies.}
\vspace*{-3ex}
\label{fig:opts-scalability}
\end{center}
\end{figure}

As is shown in Figure~\ref{fig:opts-scalability}, the scalability of the operations 
varies significantly under different thread affinity strategies. In addition, speedup 
values peak when the number of threads is a multiple of the number of computation cores 
(60, 120, 180, and 240 threads). This is expected because executing the same number of 
threads as the number of cores reduces computational imbalance among the cores. 
Morphological Open and Distance Transform achieve the best performances with 
120 and 240 threads, respectively. The performance of Morphological Open decreases after
120 threads because it is a memory-intensive operation. As is reported by other 
researchers~\cite{McCalpin1995,DBLP:conf/ipps/SauleC12}, the peak memory bandwidth of 
MIC for regular data access (STREAM benchmark~\cite{McCalpin1995}) is
attained with 1 or 2 threads per core. If more cores are
used in data intensive operations, the bandwidth decreases because of  
overload on the memory subsystem.

Distance Transform scales well until 240 threads are dispatched. The performance 
of this operation also depends on the memory bandwidth of the MIC. The 
difference is that this operation's data access pattern is irregular. To 
better understand the performance behavior of Distance Transform and because 
memory bandwidth figures for irregular data access are not provided in the 
hardware specifications, we created a micro-benchmark to measure the memory 
bandwidth of the MIC for random data accesses. The micro-benchmark program reads 
and writes data elements from/to random locations in a matrix in parallel. 
Locations to be accessed in the matrix are pre-computed and stored in a secondary 
array. Our experiments showed that the benchmark program scaled until 240 threads 
were executed. This finding correlates with the observed performance of Distance 
Transform.

The performances of the Scatter and Balanced thread affinity strategies were
similar for both operations. The only significant discrepancy among the
strategies occurred for Morphological Open and the Compact affinity strategy.
This is because Compact utilizes the 60 physical cores only when 240 threads
are dispatched. However, the Morphological Open achieves its best performance
when 120 threads are created (2-way hyperthreading) with 60 cores.

\subsection{Performance Analysis of Core Operations on MIC, GPU and CPU} \label{sec:comp-performance}
These experiments compare the speedups of core operations on MIC, GPU and CPU.
We computed the speedup values for an operation using the single-core CPU
execution as the baseline performance. We used single-core implementations from
well-known libraries and implementations published by other research groups
(see Section~\ref{sec:core-opts}). We have classified the core operations into three groups based on data access 
and computation patterns in order to better understand and characterize the 
speedup values obtained in the experiments. The groups are:
(1)~Operations with regular data access: RGB2Gray, Morphological Open, Color
Deconvolution, Pixel Stats, Gradient Stats, and Sobel Edge; (2)~Operations with
irregular data access: Morphological Reconstruction, FillHoles, and Distance
Transform; (3)~Operations that heavily use atomic instructions: Area
Threshold and Connected Component Labeling (CCL). We draw from the Roofline 
model proposed by Williams et al.~\cite{Williams:2009:RIV:1498765.1498785} 
in our evaluation. 
\subsubsection{{\bf Regular Operations.}}  
The performance of an operation in this class heavily depends on the memory 
bandwidth of a computing device. We measured the memory bandwidths of the 
GPU, CPU and MIC using the STREAM benchmark~\cite{McCalpin1995} which are 
148GB/s, 78GB/s (total bandwidth of two CPUs) and 160GB/s 
(one or two threads per core achieve similar performance), respectively. 
The peak computation capacities of the K20 GPU and MIC for double 
precision numbers are about 1~TFLOPS each, while 2 CPUs together 
achieve 345~MFLOPS.

\begin{figure*}[htb!]
\begin{center}
\includegraphics[width=\textwidth]{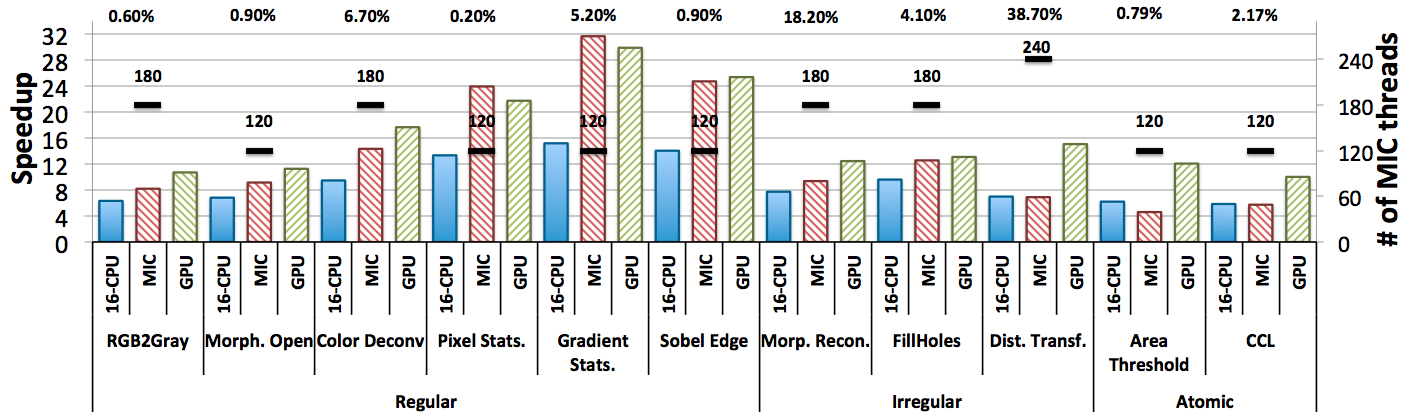}
\vspace*{-4ex}
\caption{Speedups of the core operations on multi-core CPU, MIC and GPU, 
using the single-core CPU versions as baseline. The number above each dash 
is the number of threads that led to the best performance on the MIC. The
number above the graph in each column is the percent contribution of the 
corresponding operation to the overall application execution time.}
\vspace*{-4ex}
\label{fig:comp-performance}
\end{center}
\end{figure*}

The Morphological Open, RGB2Gray and Color Deconvolution operations are memory
bound with low arithmetic-instruction-to-byte ratio. Their speedups are low on
the CPU because they quickly saturate the memory bus. Their performances on the 
GPU are nearly 1.25$\times$ higher than on the MIC as is shown in 
Figure~\ref{fig:comp-performance}. The remaining 
regular operations (Pixel Stats, Gradient Stats and Sobel Edge) are compute
bound due to their higher computation intensity. These operations achieve
better scalability on all the devices. The GPU and the MIC result in similar 
performance improvements for these operations. Their execution times improve by 
about 1.9$\times$ over the multi-core CPU 
executions. This group of compute intensive operations 
achieve the best performance on the MIC using 120 threads. Using more threads 
does not improve performance because the MIC threads can launch a vector instruction 
every two cycles. 
\subsubsection{{\bf Irregular Operations.}}
The operations with irregular data access patterns include
Morphological Reconstruction, FillHoles and Distance Transform. We measured 
the bandwidth of each device for random data accesses using the micro-benchmark 
program described in Section~\ref{sec:opts-scalability}. The program executed 
10 million random reads and writes on a 4K$\times$4K matrix of integer 
values. 
The results are presented in Table~\ref{tab:random-bandwidth}. The GPU has 
much higher bandwidth as compared to the other devices. The performance of 
the MIC is low, especially for writes. This is a result of heavy data traffic 
on the ring bus connection in the MIC which is necessary to maintain
consistency. 
\begin{table}[h!]
\caption{Device bandwidths for random data accesses (MB/s).}
\vspace*{-2ex}
\begin{center}
\begin{tabular}{l l l l}
\hline
   	& CPU	& MIC	&	GPU 	\\ \hline \hline
Reading	& 305	& 399	&	895	\\ \hline
Writing	& 74	& 16	&     	126	\\ \hline
\end{tabular}
\end{center}
\label{tab:random-bandwidth}
\vspace*{-2ex}
\end{table}

Given the micro-benchmark results, it is not surprising that 
Distance Transform is much faster (about two times faster)
on the GPU than on the MIC or CPU as is shown in 
Figure~\ref{fig:comp-performance}. All phases of this operation perform only
irregular data accesses with significantly more data reading than writing. Thus, 
the random data access bandwidths of the devices have significant impact on 
the operation's overall performance.
Morphological Reconstruction and Fill Holes, on the other hand, have different 
performance behaviors than Distance Transform, because their data access 
patterns are not purely irregular. These operations have two computation 
phases. They first perform regular raster-/anti-raster passes on data. Then, 
they carry out irregular computations that use a queue to execute propagations. 
As such, the operations benefit more from the MIC processor, which is efficient in their
regular phase. Additionally, they may execute the regular phase a
number of times before the second phase. This value was tuned in the experiments 
for each device. The regular phase was 
executed a large number of times on the MIC, before the second phase was started. 
Hence, for these operations, the MIC compares better with the GPU; the GPU is
only 1.33$\times$ faster than the MIC.
\subsubsection{Operations that Rely on Atomic Instructions.} 
Atomic instructions are necessary for correct execution in some operations; Area 
Threshold and CCL intensively use atomic instructions. We developed
another micro-benchmark to measure device bandwidths when atomic 
instructions are used. The benchmark program implements two scenarios. In the first 
scenario, a single variable is updated by all threads. This is a worst-case scenario 
because all threads access the same memory address and cause a lot of congestion on 
atomic instruction executions. In the second scenario, the elements of an array are 
updated such that an element is updated by one thread only; however, a thread may update 
multiple array elements and multiple array elements may be updated concurrently. The 
second scenario does not cause congestion on atomic instruction executions, although 
each thread executes an atomic instruction when accessing an array element. Less 
congestion is expected to yield better performance on all the devices. 
\begin{table}[h!]
\caption{Measured device bandwidths for the Atomic Add operation (Millions/sec).}
\vspace*{-2ex}
\begin{center}
\begin{tabular}{l l l l}
\hline
   		& CPU	& MIC		&	GPU 	\\ \hline \hline
Single Variable	& 134	& 55		&	693	\\ \hline
Array		& 2,200	& 906		&	38,630  \\ \hline
\end{tabular}
\end{center}
\label{tab:atomic}
\vspace*{-2ex}
\end{table}

The results of the benchmark runs are presented in Table~\ref{tab:atomic}. 
The bandwidth of the GPU is the highest in both of the scenarios. However, 
the GPU is also the
device that suffers the highest performance degradation from the second 
scenario to the first (worst-case) scenario. This large decrease in 
performance 
is a consequence of the fact that warps of threads on a GPU must
execute the same instructions. It is expected that all threads
executing in a warp will cause a conflict when updating the memory 
in the single variable case, and, consequently, threads will be 
serialized. This problem is
intensified in the GPU because it supports a much higher level of concurrency
than the other devices. The CPU is almost 2.4$\times$
faster than the MIC in both scenarios. The MIC relies on the use of
vectorized instructions for good performance, since it is equipped with
a simpler computing core. However, it does not support atomic vectorized
instructions, such as those proposed by Kumar et. al.~\cite{4556746} for other
multiprocessors. 

As is expected from the benchmark results, Area Threshold and CCL achieve higher 
performance on the GPU than on the other two devices, with executions on the CPU 
being faster than those on the MIC. 
\subsection{Cooperative Execution on Hybrid Machines}
%
% This section evaluates the cooperative use of GPUs, MICs, and CPUs to execute 
% the microscopy image analysis application.
%
\paragraph{{\bf Comparing Performances of Schedulers.}}
This experiment compares the performance of proposed and baseline schedulers.
The experiment uses 1~GPU, 1~MIC, and 14~CPU cores for
computation; the remaining 2~CPU cores are used to manage the co-processors. The
input data contains 800 4K$\times$4K image tiles. Because the decision space
may affect the performances of the schedulers, we varied the number of
coarse-grained application stages that were concurrently executed in a Worker. This value 
is referred to as the {\em window size}. The window size was varied from 16 (the number of
computing slots) to 70 -- values larger than 70 had no significant effect 
on performance. The
speedup and execution time estimates used by the schedulers were determined from 
previous runs. 
%We discuss the impact of the 
%accuracy of these estimates on scheduler performance in 
%Section~\ref{sec:sensibility}. 
%
\begin{figure}[h!]
	\centering
	\includegraphics[width=0.8\linewidth]{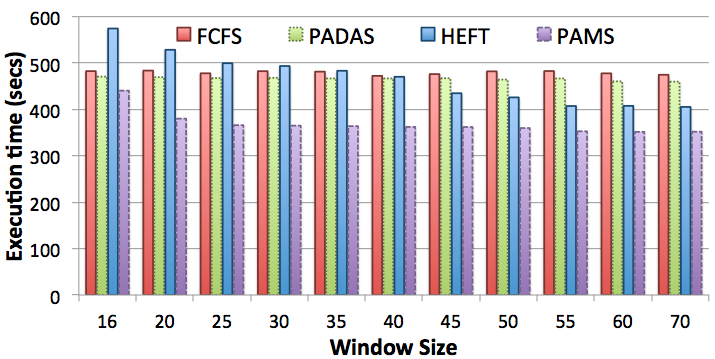}
\vspace{-2mm}
	\caption{Performance of the scheduling strategies when the number of 
	concurrent tasks assigned to a Worker (window size) is varied.}
	\label{fig:windowVariation}
\end{figure}

The execution times of the image analysis application using FCFS, PADAS, HEFT and 
PAMS are shown in Figure~\ref{fig:windowVariation}. FCFS and PADAS result in similar
execution times. Window size has little impact on their
performance. HEFT results in the highest execution times with
small window sizes, but performs better than 
FCFS and PADAS with window sizes larger than 45. HEFT performs better with larger
decision spaces because it is able to balance well the work among processors and
minimize miss-assignment of tasks.
The PAMS scheduler achieves the best performance among all the schedulers 
for all window sizes. It is at least 1.15$\times$ faster
than HEFT, but its gains are up to 1.39$\times$ for configurations with small
window sizes. This scheduler is able to perform very well even with small
window sizes and is less sensitive to window size. This is an important attribute
of a scheduler, because smaller window sizes in Workers 
reduces load imbalance among nodes on a distributed-memory machine.

%\paragraph{{\bf Assessing Performance of Schedulers.}}
%
To better understand the relative performance of the schedulers, we looked at the 
task scheduling computed by a scheduler with respect to the speedups of the operations 
on the devices and the percent 
contribution of the operations in the overall application execution time 
(Figure~\ref{fig:comp-performance}). 
\begin{figure*}[htb!]
\centering
\mbox{
    \subfigure[FCFS]{
    \includegraphics[width=0.45\linewidth]{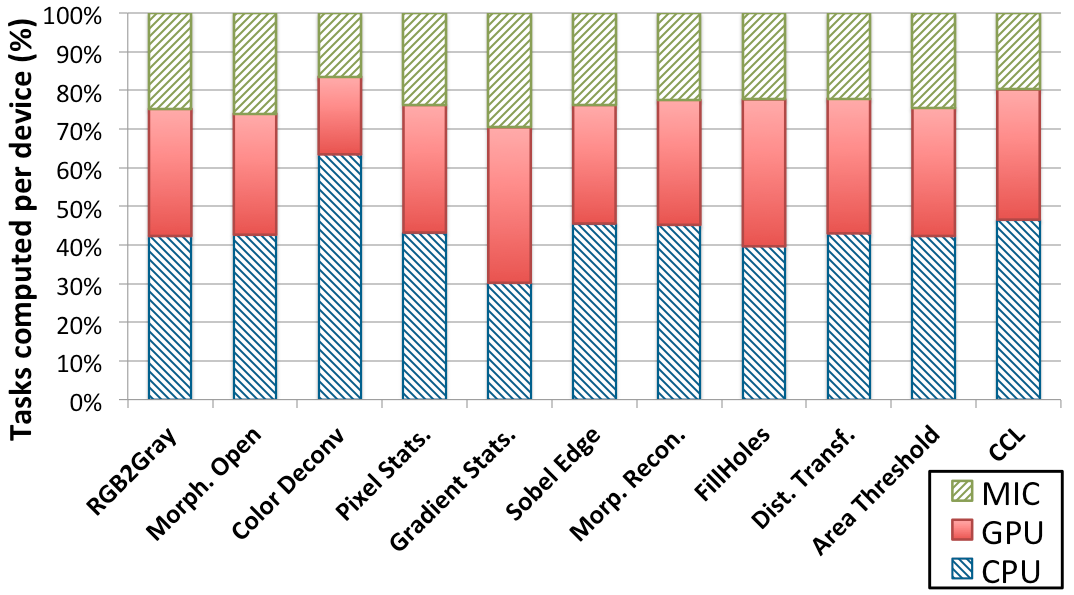}}
}
\mbox{
    \subfigure[PADAS]{
    \includegraphics[width=0.45\linewidth]{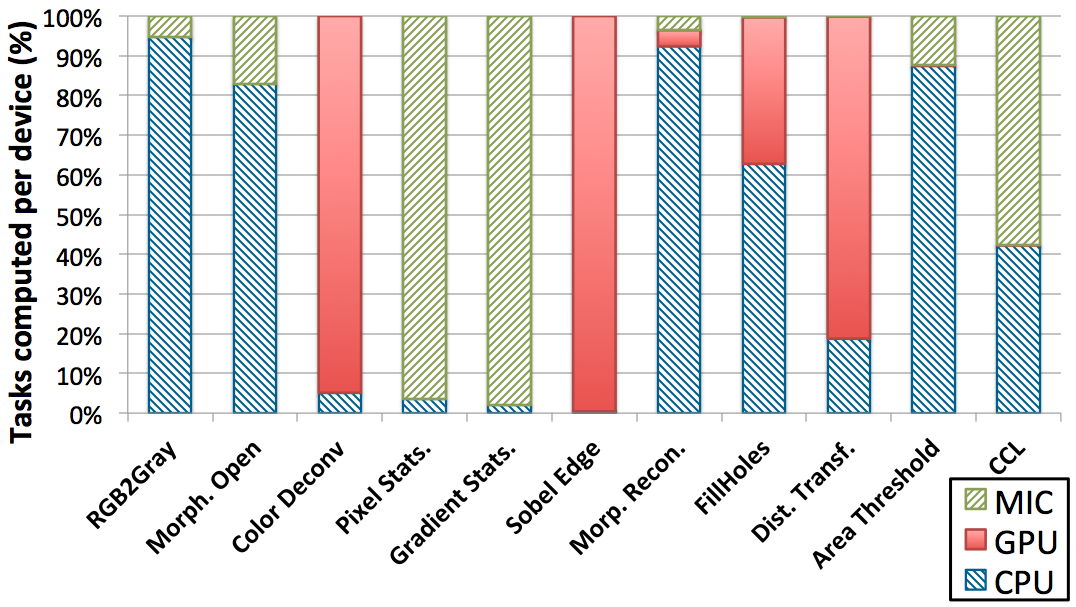}}
}
\mbox{
    \subfigure[HEFT]{
	\label{fig:prof_heft}
    \includegraphics[width=0.45\linewidth]{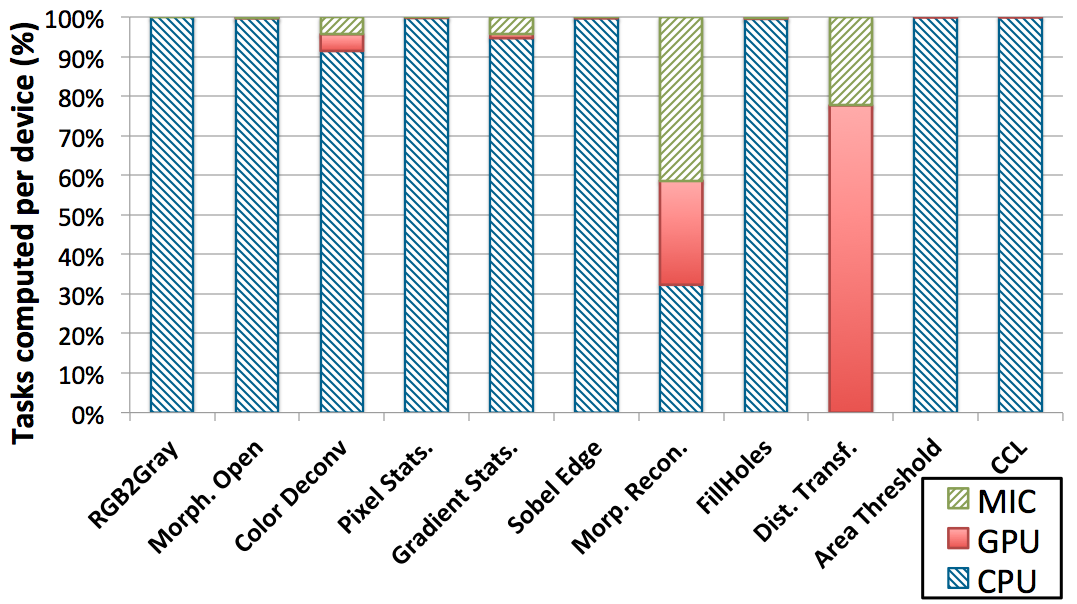}}
}
\mbox{
    \subfigure[PAMS]{
    \label{fig:pams}
    \includegraphics[width=0.45\linewidth]{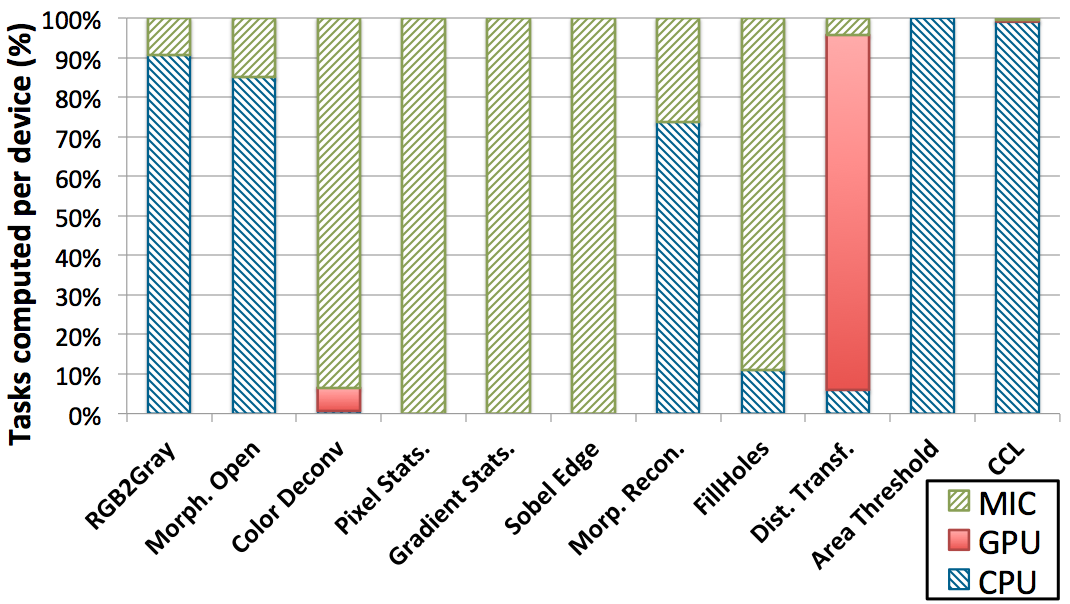}}
}
\vspace{-1ex}
\caption{Task assignment profile for each of the operations executed in the image 
analysis application under different scheduling strategies. A fixed window size of 
$80$ is used in all runs.} 
\vspace{-1ex}
\label{graf:distrib}
\end{figure*}
The percentage of operations executed by each computing device for a given 
scheduling strategy is presented in Figure~\ref{graf:distrib}. The number of
operations that FCFS assigns to each device is almost the same across 
all the operations. PADAS computes a
scheduling in which operations are preferably assigned to certain types of
processing devices. Nevertheless, it is not able to significantly improve on 
the performance of FCFS. 
PADAS fails when 
it assigns a substantial number of
Distance Transform tasks to CPU cores although Distance Transform is more efficient 
on co-processors.
The Sobel operation illustrates well what we believe to be the major problem with
this policy. Sobel achieves the highest speedups on the GPU, but it is also efficient on
the MIC. Because it is faster on the GPU, it is inserted in the queue from the 
GPU end. This significantly reduces the chances of the MIC getting 
selected for this operation. As a consequence, the MIC is used with other
operations for which it is less efficient.

The analysis of the HEFT scheduling (Figure~\ref{fig:prof_heft}) shows that
most of the tasks with short execution times were executed by the CPU cores, 
regardless of the speedup values. HEFT also assigns all of the
Distance Transform operations to the co-processors, but this operation is
2$\times$ more efficient on the GPU than on the MIC. Since no other device is able to
efficiently perform this computation, the GPU should be used with higher
percentages than what was observed in the experiments. Moreover, 
because this operation is
responsible for nearly 40\% of the whole application execution time, its
correct scheduling is critical for high performance. 

The profile of PAMS scheduling shows a better distribution of tasks. It 
schedules most of the Distance Transform operations on the GPU and 
utilizes the CPU cores for operations (such as Area Threshold and CCL) 
that do not attain high performance on the co-processors. It also schedules 
operations (such as ColorDevonv, FillHoles, Gradient, Pixel Stats and Sobel 
Edge) to the MIC that benefit from this device.
\paragraph{{\bf Evaluating Different Configurations of Hybrid System.}}
This section investigates how well the scheduling strategies perform when 
different combinations of CPU and co-processors are used as the hybrid 
system. In this work we looked at three hardware configurations: CPU-GPU 
(15 CPU cores and 1 GPU), CPU-MIC (15 CPU cores and 1 MIC) and CPU-GPU-MIC 
(14 CPU cores , 1 GPU, and 1 MIC). 
\begin{figure}[htb!]
	\centering
	\includegraphics[width=0.6\linewidth]{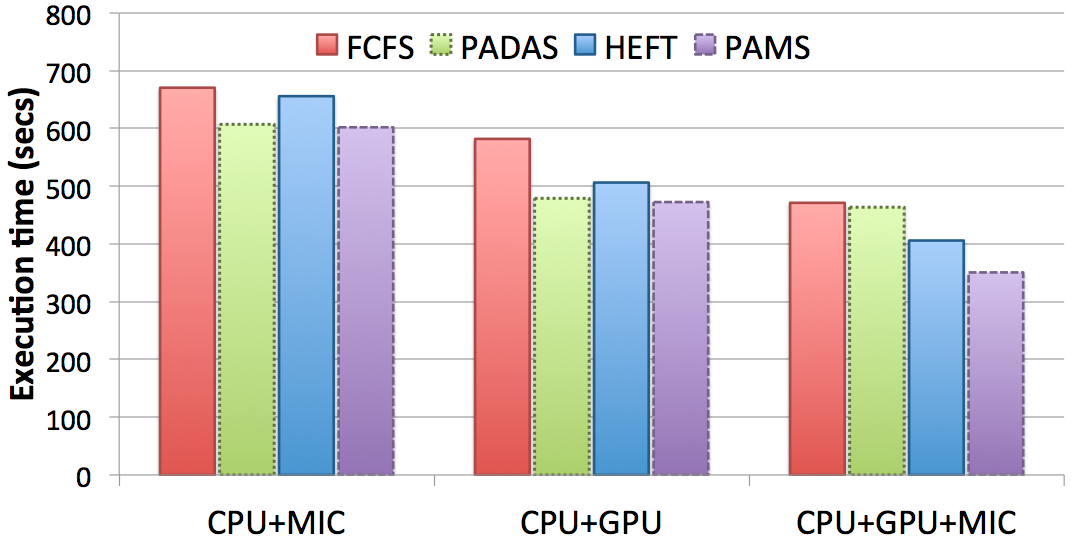}
	\vspace{-2mm}
	\caption{Application performance when application tasks are scheduled and 
	mapped to two or more computation devices.}
	\label{fig:procVariation}
	\vspace{-1mm}
\end{figure}

The experimental results presented in Figure~\ref{fig:procVariation} show
that FCFS has the worst performance in all hardware configurations. In the
scenarios with two device types (CPU-MIC and CPU-GPU), PADAS and PAMS
result in nearly the same application execution times. This is expected because these
schedulers will compute the same scheduling, since PAMS is reduced to PADAS
with an implementation using two queues instead of one. PADAS and PAMS
outperform HEFT in all of the configurations. The application executes faster in 
the configuration with three devices with PAMS. 
The configuration
with CPU-GPU is about 1.27$\times$ faster than the one with CPU-MIC. 
This is because of the fact that neither CPU nor MIC is efficient with
irregular patterns and atomic operations, which account for a large amount
of the application execution time. 
\paragraph{{\bf Sensitivity to Inaccurate Estimates.}} \label{sec:sensibility}
This section empirically evaluates the ability of the schedulers to perform well
in cases in which estimations of the metrics (execution time and speedup) they 
use in computing a schedule are inaccurate. To carry out these experiments, we
introduced errors in the execution times and, as a consequence, in the
speedups in a controlled manner. The amount of error was varied as 
$0\%$, $10\%$, $25\%$, $50\%$, $75\%$ and $100\%$ of the execution
times of the operations, with equal probability that the error was an increase 
or a decrease in execution time. 
\begin{figure}[htb!]
	\centering
	\includegraphics[width=0.8\linewidth]{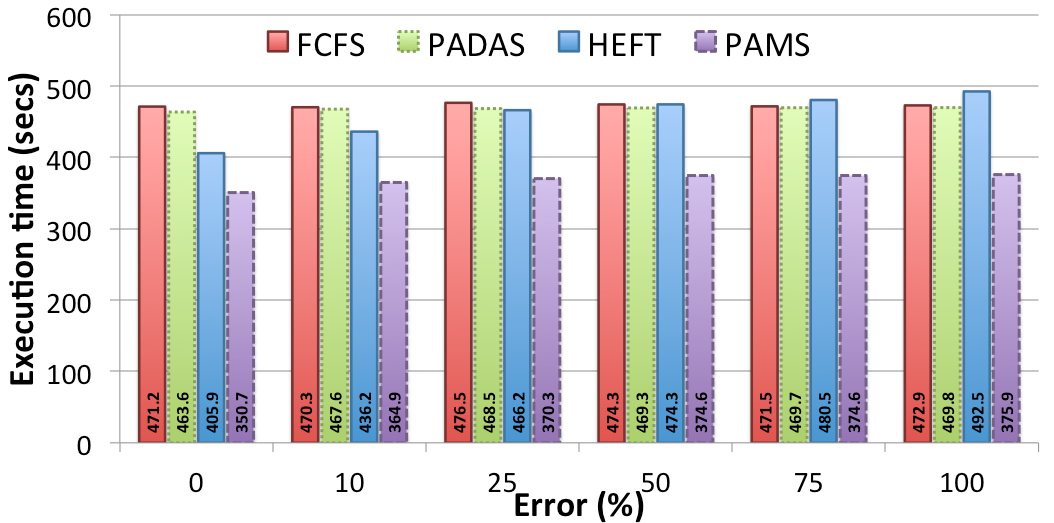}
	\vspace{-2mm}
	\caption{Performance of scheduling strategies as errors are 
	introduced in estimated execution times of operations.}
	\label{fig:errorVariation}
	\vspace{-2mm}
\end{figure}

\begin{figure}[h!]
	\centering
	\includegraphics[width=0.7\linewidth]{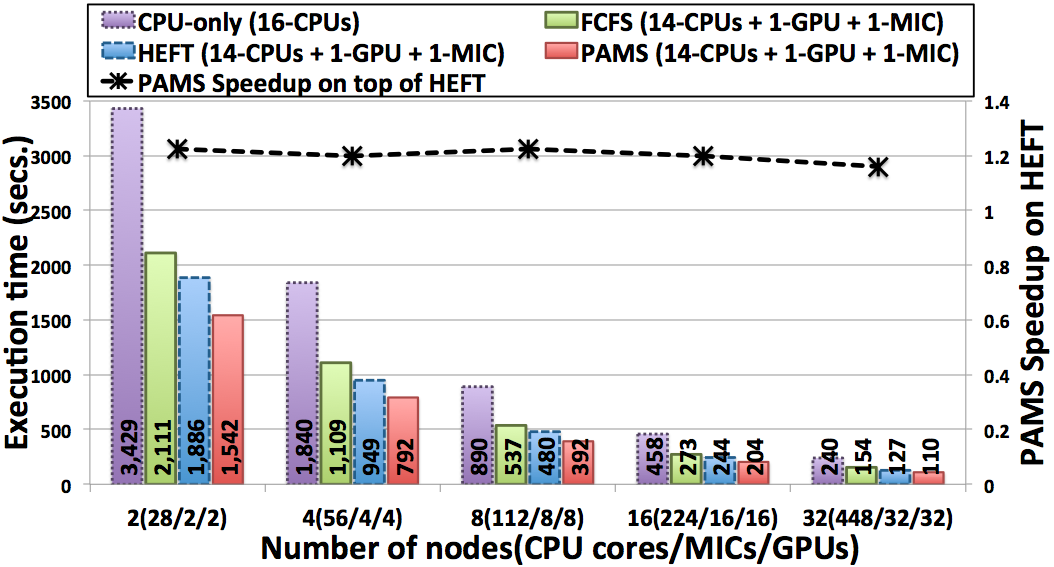}
	\vspace{-1ex}
	\caption{Strong-scaling results on a distributed-memory cluster in which each 
	node has two 8-core CPUs, one MIC, and one GPU.}
	\label{fig:scalability}
	\vspace{-2ex}
\end{figure}

As presented in Figure~\ref{fig:errorVariation}, the performances of PADAS and
PAMS are not significantly affected as the amount of error is varied. On the other
hand, HEFT, which uses execution times to calculate schedules, is strongly
impacted by the errors and is even less efficient than FCFS when the amount of 
error is higher 
than 50\%. The experimental results show that the scheduling
strategies based on the speedup metric are much less sensitive to inaccurate 
estimates and perform well even with large error in estimates. 
These schedulers use the input metric only to order tasks in a queue. The amount 
of error in estimated speedup has to be large enough to significantly alter the 
positions of tasks in the queue in order to adversely impact computed schedules 
and overall application performance. 
\paragraph{{\bf Scalability Results.}}
These experiments assessed the performance of the image analysis 
application on a distributed memory cluster. 
%The application is built from the core
%image analysis operations (Figure~\ref{fig:dataflow} in
%Section~\ref{sec:app}) and is expressed as a two-level hierarchical pipeline.
%The first level consists of segmentation and feature computation stages. Each 
%of these stages is composed of operations shown in Figure~\ref{fig:dataflow} which
%form the second level pipelines. 
The scheduling strategies used the speedup values 
obtained in the experiments in Section~\ref{sec:comp-performance} (see 
Figure~\ref{fig:comp-performance}). 
Figure~\ref{fig:scalability} presents the performance results of both the
CPU-only version of the application (16-CPUs), which uses only the 16 CPU 
cores on each node, and the hybrid version, which uses the CPU, GPU and MIC 
in a coordinated manner via FCFS, HEFT or PAMS. The input data for this 
experiment consists of 6,379 4K$\times$4K image tiles. Both versions of the 
application achieved good scalability with efficiency higher than 90\%. 
The hybrid version with PAMS results in a performance improvement of 
about 2.2$\times$ on top of the CPU-only version. The performance 
improvement by PAMS on top of HEFT is maintained as the 
application scales.

We also executed the application on a large number of computing nodes equipped 
with two CPUs and one MIC only, since Stampede has more nodes with this configuration.
We experimented with five versions of the application:
CPU-only refers to the multi-core CPU version that uses all the 
CPU cores available but no co-processors; ~MIC-only uses only the MIC 
in each node to perform computations;
CPU-MIC uses all the CPU cores and the MIC in coordination and 
distributes tasks to devices in each node via FCFS, PAMS, or HEFT as
the scheduling strategy.  
\begin{figure}[htb!]
\begin{center}
	\includegraphics[width=0.7\textwidth]{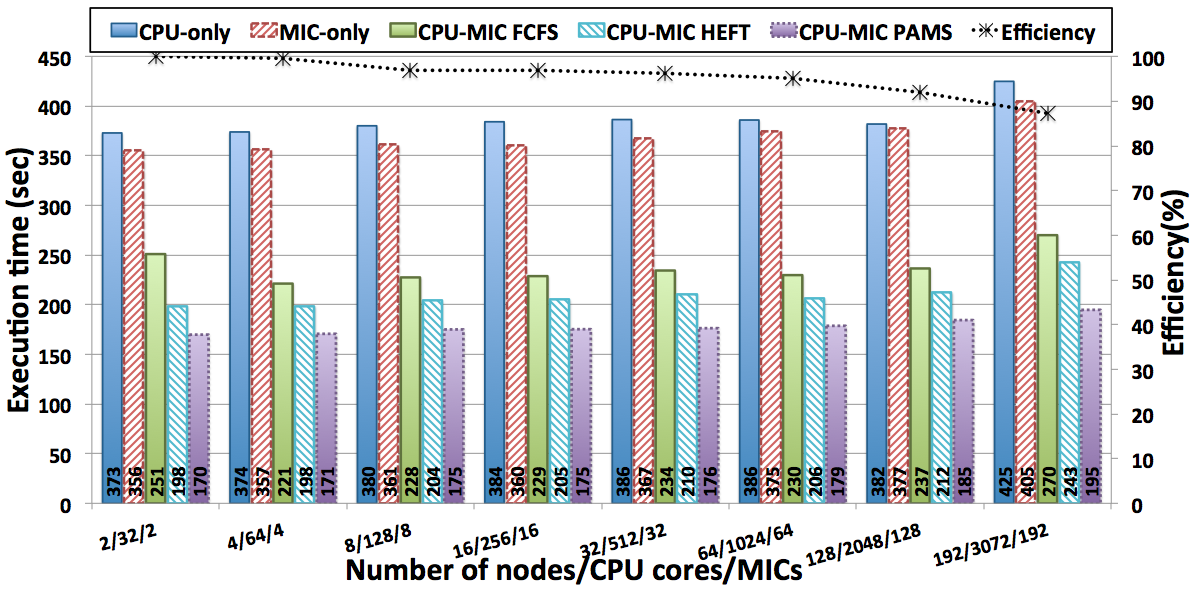}%}
	\vspace*{-1ex}
	\caption{Weak scaling results on a distributed-memory cluster in which each node 
	has 	two 8-core CPUs and one MIC.}
	\vspace*{-2ex}
\label{fig:weak-scale-general}
\end{center}
\end{figure}

Figure~\ref{fig:weak-scale-general} shows the weak scaling performance of 
the application when the
dataset size and the number of nodes increase proportionally. For the
configuration with 192 nodes, the input data contains 68,284 4K$\times$4K image
tiles for a total of 3.27~TB of uncompressed data. The application
scaled well with about 84\% of efficiency on 192 nodes. The main factor limiting scalability is the
increasing cost of reading the input image tiles concurrently from storage as the
number of nodes (and processes) grows. The CPU-MIC versions resulted in improvements of 
up to 2.06$\times$ on top of the MIC-only version. 
The CPU-MIC PAMS version results in the lowest execution times and is 
on average 1.29$\times$ faster than the CPU-MIC FCFS version.

\section{Related Work} \label{sec:related}
Cooperative use of CPU and GPU has been target of a number of
projects~\cite{mars,merge,qilin09luk,1616864,Diamos:2008:HEM:1383422.1383447.Harmony,6061070,ravi2010compiler,6152715,hpdc10george,6267858,Rossbach:2011:POS:2043556.2043579,Teodoro:2012:Cluster,augonnet:hal-00725477,cluster09george,HartleySC10,hartley,teodoro-2013-vldb}.
The Mars~\cite{mars} and Merge~\cite{merge} systems are seminal works, which
target the acceleration of MapReduce applications. Mars has evaluated the
static partition strategies for dividing MapReduce tasks between CPU and GPU.
Merge, on the other hand, proposed the use of a dynamic work partition scheme
during application execution.  Qilin~\cite{qilin09luk} presented an automatic
approach for distributing tasks for computation in hybrid systems.
PTask~\cite{Rossbach:2011:POS:2043556.2043579} provides OS abstractions for
task based applications on GPU equipped systems.  These systems are designed
for shared memory machines.

Runtime systems for execution on distributed memory systems have also been
developed~\cite{6061070,ravi2010compiler,HartleySC10,hpdc10george,6267858,Teodoro:2012:Cluster,augonnet:hal-00725477,Teodoro2014589,andrade2014efficient}.
DAGuE~\cite{6061070} and StarPU~\cite{1616864,augonnet:hal-00725477} have been
successfully applied in regular linear algebra applications. These systems
describe applications as directed acyclic graphs of operations and implement
multiple scheduling policies, e.g., including those that prioritize computation
of critical paths in the dependency graph in order to maximize parallelism.
Ravi~\cite{ravi2010compiler} and Hartley~\cite{HartleySC10} proposed runtime
techniques to auto-tune work partitioning among CPUs and GPUs.
OmpSs~\cite{6267858} supports execution of dataflow applications created via
compilation of annotated codes. The performance-aware scheduling strategies
employ proposed in this work use a different priority metric (relative
performance or speedup) of operations on CPUs or accelerators for mapping of
work to processors. As presented, it results into significant performance
improvement on top of HEFT, which is widely used in previous works and is known
as a very efficient scheduling policy.

The release of the new Intel Xeon Phi has motivated a number of
research efforts~\cite{phi-1,6569806,phi-3,phi-4,DBLP:conf/ipps/SauleC12}. Hamidouche
et al.\cite{Hamidouche:2013:MEE:2464996.2465445} have developed a very
interesting platform that allows for off-loading computations to a MIC on
a remote machine. In this way, applications running on multiple nodes may share 
a MIC to maximize its utilization. The use of OpenMP to develop code
targeting MICs has also been studied~\cite{phi-3}. The authors 
analyzed the overheads of important OpenMP directives for thread creation and 
synchronization. They also evaluated the processor performance with respect
to memory bandwidth.  Saule et al.~\cite{DBLP:conf/ipps/SauleC12} implemented
optimized sparse matrix multiplication kernels for MICs and provided a
comparison of MICs and GPUs for this operation.
 
In our work, we perform a comparative performance evaluation of MICs,
multi-core CPUs, and GPUs using an important class of operations.  These
operations employ diverse computation and data access patterns and several
parallelization strategies. The comparative performance analysis correlates the
performance of operations with co-processors characteristics using co-processor
specifications or performance measured using micro-kernels. This evaluation
provides a methodology and insights towards a better understanding of the
efficacy of co-processors for a class of applications. We also
investigate coordinated use of MICs, GPUs and CPUs on a distributed memory machine
and its impact on application performance. Our approach takes into account
performance variability of operations to make smart task assignments.

\section{Conclusions} \label{sec:conclusions}
Development of application optimizations to maximize performance benefits from hybrid
systems is a very challenging problem. Besides optimizations for load balancing and 
minimizing data transfers between devices, device hardware and execution 
characteristics need to be taken into account.
With increasing availability of 
high-end machines with multiple co-processors with different
architectures, application developers have to
understand the effects of different devices on the performance of their applications. 
However, little information about the relative
performance of modern co-processors for different computation patterns is
available. 

In this work, we aim to address this issue by conducting a systematic performance 
comparison of CPUs, GPUs and MICs using a set of operations with diverse data access 
patterns (regular and
irregular), computation intensity, and types of parallelization
strategies from an important class of applications.  We also propose
and evaluate new performance-aware scheduling techniques for efficiently
executing complex applications in systems equipped with CPUs and multiple
accelerators. Our scheduling techniques take into account performance
variabilities across computation tasks or operations to better utilize hybrid
systems. 

The experimental results show that different types of co-processors are more
appropriate for specific data access patterns and types of parallelism, as
expected. The MIC's performance compares well with that of the GPU when regular
operations and computation patterns are used. However, the GPU is significantly
more efficient for operations that perform irregular data access and heavily
employ atomic operations. Our work also shows that a strong performance
variability exists among different operations, as a result of their computation
patterns. This performance variability needs to be taken into account to
efficiently execute pipelines of operations using co-processors and CPUs in
coordination. 

We compare the proposed scheduling strategies (PADAS and PAMS) with two well known 
scheduling policies: FCFS and HEFT on a distributed-memory cluster of nodes with 
multiple types of co-processors. The results show that PAMS achieves better performance than other
polices, being at least 1.15$\times$ faster than its closest competitor: HEFT.
We also experimentally show that performance-aware strategies perform well even
when a high level of inaccuracy exists in the input information (speedups) used
to compute the scheduling. On the other hand, the HEFT time based scheduler
suffers dramatically as the estimation error in the execution times of
operations increases. The example application achieves good scalability in
distributed memory settings via co-operative use of CPUs and 
co-processors. \\

\noindent {\bf Acknowledgments.}
{\small
This work was supported in part by HHSN261200800001E and 1U24CA180924-01A1
from the NCI, R24HL085343 from the NHLBI, R01LM011119-01 and
R01LM009239 from the NLM, RC4MD005964 from the NIH, National Institutes of
Health (NIH) K25CA181503 and by CNPq,
CAPES, FINEP, Fapemig, and INWEB. This research used resources provided by the XSEDE
Science Gateways program and the Keeneland Computing Facility at the Georgia
Institute of Technology, which is supported by the NSF under Contract
OCI-0910735.}

\bibliographystyle{vancouver}
\bibliography{george}

\end{document}